\documentclass[conference]{IEEEtran}

\usepackage{times,graphics,amssymb,array,amsmath,bm,cite}
\usepackage{algorithmic}
\usepackage{algorithm}
\usepackage[dvips]{graphicx}
\usepackage[active]{srcltx}  
\usepackage{epsfig,latexsym}
\usepackage{booktabs}
\usepackage{glossaries}
\usepackage{hyperref}

\usepackage[dvipsnames,svgnames]{xcolor}
\usepackage[textwidth=30mm]{todonotes}

\newcommand{\eeq}{\end{equation}}

\newcommand{\ds}{\displaystyle}

\newcommand{\beq}{\begin{equation}}

\newcommand{\tr}      {{\mathrm{tr}}}    

\newtheorem{theorem}{Theorem}

\makeglossaries

\newacronym{mac}{MAC}{multiple-access channel}
\newacronym{bc}{BC}{broadcast channel}
\newacronym{mimo}{MIMO}{multiple-input multiple-output}
\newacronym{siso}{SISO}{single-input single-output}
\newacronym{sc}{SC}{single-carrier}
\newacronym{mc}{MC}{multi-carrier}
\newacronym{ofdma}{OFDMA}{orthogonal frequency division multiple access}
\newacronym{af}{AF}{amplify-and-forward}
\newacronym{df}{DF}{decode-and-forward}
\newacronym{cf}{CF}{compress-and-forward}
\newacronym{mwrc}{MWRC}{multi-way relay channel}
\newacronym{pde}{PDE}{partial data exchange}
\newacronym{fde}{FDE}{full data exchange}
\newacronym{iid}{i.i.d.\@}{independent and identically distributed}
\newacronym{awgn}{AWGN}{additive white Gaussian noise}
\newacronym{awg}{AWG}{additive white Gaussian}
\newacronym{sic}{SIC}{successive interference cancellation}
\newacronym{snr}{SNR}{signal-to-noise ratio}
\newacronym{sinr}{SINR}{signal to interference plus noise ratio}
\newacronym{ber}{BER}{bit error rate}
\newacronym{zf}{ZF}{zero-forcing}
\newacronym{mmse}{MMSE}{minimum mean square error}
\newacronym{sud}{SUD}{single user decoding}
\newacronym{dof}{DoF}{degrees of freedom}
\newacronym{gdof}{GDoF}{generalized degrees of freedom}
\newacronym{nnc}{NNC}{noisy network coding}
\newacronym{dmn}{DMN}{discrete memoryless network}
\newacronym{csi}{CSI}{channel state information}
\newacronym{ee}{EE}{energy efficiency}
\newacronym{ian}{IAN}{treating interference as noise}
\newacronym{snd}{SND}{simultaneous non-unique decoding}
\newacronym{brd}{BRD}{best response dynamics}
\newacronym{br}{BR}{best response}
\newacronym{ne}{NE}{Nash equilibrium}
\newacronym{gne}{GNE}{generalized Nash equilibrium}
\newacronym{lhs}{LHS}{left-hand side}
\newacronym{rhs}{RHS}{right-hand side}
\newacronym{gee}{GEE}{global energy efficiency}
\newacronym{wsee}{WSEE}{weighted sum energy efficiency}
\newacronym{wpee}{WPEE}{weighted product energy efficiency}
\newacronym{wmee}{WMEE}{weighted minimum energy efficiency}
\newacronym{kkt}{KKT}{Karush Kuhn Tucker}
\newacronym{pc}{PC}{pseudo-concave}
\newacronym{qc}{QC}{quasi-concave}
\newacronym{ql}{QL}{quasi-linear}
\newacronym{pl}{PL}{pseudo-linear}
\newacronym{spc}{SPC}{strictly pseudo-concave}
\newacronym{sqc}{SQC}{strictly quasi-concave}
\newacronym{lfp}{LFP}{linear fractional problem}
\newacronym{clfp}{CLFP}{concave-linear fractional problem}
\newacronym{ccfp}{CCFP}{concave-convex fractional problem}
\newacronym{mmfp}{MMFP}{max-min fractional problem}
\newacronym{sorp}{SoRP}{sum-of-ratios problem}
\newacronym{porp}{PoRP}{product-of-ratios problem}
\newacronym{srp}{SRP}{single-ratio problem}
\newacronym{brb}{BRB}{branch-reduce-and-bound}
\newacronym{qos}{QoS}{quality-of-service}
\newacronym{comp}{CoMP}{cooperative multi-point}
\newacronym{ue}{UE}{user equipment}
\newacronym{bs}{BS}{base station}
\newacronym{mrc}{MRC}{maximum ratio combining}
\newacronym{d2d}{D2D}{device-to-device}
\newacronym{lmmse}{LMMSE}{linear minimum mean square error}
\newacronym{svd}{SVD}{singular value decomposition}
\newacronym{evd}{EVD}{eigen value decomposition}
\newacronym{ict}{ICT}{information communication technologies}
\newacronym{ann}{ANN}{artificial neural network}

\def\CN{\mathcal{N}_{\mathbb{C}}} 
\newcommand{\vect}[1]{\mathbf{#1}}
\def\tr{\mathrm{tr}}

\def\Psiv{\vect{Q}}

\def\tr{\mathrm{tr}}

\def\Htran{\mbox{\tiny $\mathrm{H}$}}
\def\Ttran{\mbox{\tiny $\mathrm{T}$}}
\def\CN{\mathcal{N}_{\mathbb{C}}} 
\def\taupu{\tau_{p}} 
\def\bphiu{\boldsymbol{\phi}} 

\def\Pu{\mathcal{P}} 

\IEEEoverridecommandlockouts

\begin{document}

\title{User Association and Load Balancing\\ for Massive MIMO through Deep Learning}
\author{
\IEEEauthorblockN{Alessio Zappone\IEEEauthorrefmark{3}, Luca Sanguinetti\IEEEauthorrefmark{2}\IEEEauthorrefmark{3},
Merouane Debbah\IEEEauthorrefmark{3}\IEEEauthorrefmark{4}\bigskip
\thanks{\hrulefill \newline The research of A. Zappone was supported by the H2020 MSCA IF BESMART, grant 749336. The research of L. Sanguinetti was partly supported by the H2020-ERC PoC-CacheMire project, grant 727682, and by the University of Pisa under the PRA 2018-2019 Research Project CONCEPT. The work of M. Debbah was partly supported by the H2020 MSCA IF BESMART, grant 749336, and by the H2020-ERC PoC-CacheMire project, grant 727682.}
}
\IEEEauthorblockA{\IEEEauthorrefmark{3}\small{Large Networks and System Group (LANEAS), CentraleSup\'elec, Universit\'e Paris-Saclay, Gif-sur-Yvette, France}}
\IEEEauthorblockA{\IEEEauthorrefmark{2}\small{Dipartimento di Ingegneria dell'Informazione, University of Pisa, Pisa, Italy}}
\IEEEauthorblockA{\IEEEauthorrefmark{4}\small{Mathematical and Algorithmic Sciences Lab, Huawei Technologies, France.}}
}
\maketitle

\begin{abstract}
This work investigates the use of deep learning to perform user-cell association for sum-rate maximization  in Massive MIMO networks. It is shown how a deep neural network can be trained to approach the optimal association rule with a much more limited computational complexity, thus enabling to update the association rule in real-time, on the basis of the mobility patterns of users. In particular, the proposed neural network design requires as input only the users' geographical positions. Numerical results show that it guarantees the same performance of traditional optimization-oriented methods. 
\end{abstract}

\section{Introduction}\label{Sec:Intro}
5G wireless networks are scheduled to be rolled-out in only a couple of years. The envisioned need of serving a very high number of devices makes it urgent to increase the data rate by a factor 1000 \cite{5GNGMN}. Massive MIMO is considered as a key 5G technology to achieve this ambitious goal \cite{marzetta2010noncooperative, BHS18,EmilNowPub17, Marzetta2016a}, thanks to its ability to simultaneously serve many users by means of spatial multiplexing. 

Several recent contributions have addressed the issue of optimally determining the association between mobile users and \glspl{bs} in multicell heterogeneous networks. In \cite{Liu2015}, the user-cell association is optimized in Massive MIMO heterogeneous networks with the aim of maximizing the network sum-utility, ensuring fairness among the users. Sum-utility maximization through user-cell assignment is also performed in \cite{Bethanabhotla16,Ye2016}, where both centralized and distributed solutions are developed. In 
\cite{Chien2016}, the problem of joint power and assignment allocation for power minimization subject to rate constraints is analyzed. It is shown that the problem can be cast as a linear program solvable with relatively low complexity. The problem of user association is addressed in \cite{Zhu2016} in the context of Massive MIMO systems powered by wireless power transfer. User association and power control are optimized in \cite{Hao2016} to achieve a trade-off between spectral and energy efficiency in Massive MIMO heterogeneous networks. Joint transmit power and assignment design is also performed in 
\cite{Lin2017}, following a proportional fairness approach. User-cell association is optimized in \cite{Zhou2017} with the goal of maximizing the sum energy efficiency of Massive MIMO heterogeneous networks. In \cite{Ma2017}, the goal is to maximize the system sum-rate, and the user-cell assignment is optimized in Massive MIMO using nested antenna arrays. In \cite{Feng2018}, the  impact of backhaul capacity constraints on the association problem is analyzed. In \cite{YuDL2018} a convolutional network is used to estimate the users' positions and perform user-cell assignment.

A common feature of all the aforementioned solutions is that they require to recompute the optimal association rule anytime the propagation scenario changes, either following fast-fading channel realizations, or, more realistically, focusing on the slow-fading effects such as users' mobility and/or shadowing. However, even focusing on slow-fading channel realizations, it is still necessary to update the optimal assignment whenever the slow-fading channel characteristics vary. In large cellular networks with many users, as envisioned for Massive MIMO networks, this entails a considerable complexity overhead, ultimately impairing a true real-time user-cell assignment. 

Motivated by this, the main contribution of this work is to show that the user-cell association can be performed in real-time in realistic Massive MIMO networks, by using a deep learning framework. Particularly, we show how deep \glspl{ann} can be trained to compute the optimal user-cell association based only on the mobile users' positions in the service area, and how this can be accomplished with much lower complexity than that required by traditional user-cell association approaches. Numerical results are provided to  confirm the merits of the proposed method in a realistic Massive MIMO network setup. 

\section{System model}\label{Sec:SystemModel}
Consider the uplink of a Massive MIMO network composed of $M$ cells, the BS of each cell comprising $N$ antennas to communicate with $K$ single-antenna UEs. We call ${\bf h}_{lk}^j \in \mathbb{C}^{N}$ the channel from BS $j$ to UE $k$ in cell $l$ within a coherence block and assume that \cite{EmilNowPub17}
\begin{equation} \label{eq:correlated-Rayleigh-model}
{\bf h}_{lk}^j\sim \CN \left( \vect{0}_{N}, {\bf R}_{lk}^j  \right)
\end{equation}
where ${\bf R}_{lk}^j \in \mathbb{C}^{N \times N}$ is the spatial correlation matrix, assumed to be known at the BS. The Gaussian distribution models the small-scale fading variations, while ${\bf R}_{lk}^j $ describes the macroscopic propagation characteristics.
The normalized trace $\beta_{lk}^j = \frac{1}{M} \tr ( {\bf R}_{lk}^j)$
is the average channel gain from an antenna at BS~$j$ to UE~$k$ in cell~$l$. Uncorrelated Rayleigh fading with ${\bf R}_{jlk}=\beta_{jlk}\vect{I}_N$ is a special case of this model, but ${\bf R}_{jlk}$ is in general
not diagonal \cite{BHS18,EmilNowPub17}.

\subsection{Channel estimation}

We assume that the BS and UEs are perfectly synchronized and operate according to a time-division duplex (TDD) protocol where UL reception is preceded by a pilot-based channel estimation phase with pilot sequences of length $\tau_p$. We assume that each cell is associated with $K$ orthogonal sequences. The pilot associated with UE $k$ in cell $j$ is denoted by $\bphiu_{jk} \in \mathbb{C}^{\tau_p}$ and has unit norm $\bphiu_{jk}^{\Htran}\bphiu_{jk} = 1$. Channel estimates at BS~$j$ are obtained from the received pilot signal $\vect{Y}_j^{p} \in \mathbb{C}^{M \times \taupu}$, given by \cite{EmilNowPub17}
\begin{align} \label{eq:uplink-pilot-model}
\vect{Y}_j^{p} = \underbrace{ \sum_{i=1}^{K} \sqrt{p}\, \vect{h}_{ji}^{j} \bphiu_{ji}^{\Ttran}  }_{\textrm{Desired pilots}} + \underbrace{\sum_{l=1,l \neq j}^{M} \sum_{i=1}^{K}  \sqrt{p} \vect{h}_{li}^{j} \bphiu_{li}^{\Ttran}  }_{\textrm{Inter-cell pilots}} + \underbrace{ \vphantom{\sum_{l=1,l \neq j}^{M} } \vect{N}_{j}^{p}}_{\textrm{Noise}}
\end{align}
where $p$ is the transmit power and $\vect{N}_{j}^{p} \in \mathbb{C}^{M \times \taupu}$ is thermal noise with i.i.d.\ elements distributed as $\CN(0,\sigma^{2})$. If the statistics are known, the minimum mean-squared error (MMSE) estimator of $\vect{h}_{li}^{j}$ can be computed as follows.
\begin{theorem} \label{theorem:MMSE-estimate_h_jli}
The MMSE estimate of $\vect{h}_{li}^{j}$ is
\begin{equation} \label{eq:MMSEestimator_h_jli}
\hat{\vect{h}}_{li}^{j}  = \vect{R}_{li}^{j} \big(\Psiv_{li}^{j}\big)^{-1}  \left(\frac{1}{\tau_p\sqrt{p}}\vect{Y}_j^{p}\bphiu_{li}\right)
\end{equation}
where $\vect{Y}_j^{p}$ is given in \eqref{eq:uplink-pilot-model} and 
\begin{equation} \label{eq:Psiv-definition}
\Psiv_{li}^{j} =  \sum_{l' \in \Pu_{l} }\vect{R}_{l'i}^{j}  +  \frac{1}{\tau_p}\frac{\sigma^2}{p}  \vect{I}_{M}.
\end{equation}
The estimation error $\tilde{\vect{h}}_{li}^{j} = \vect{h}_{li}^{j} - \hat{\vect{h}}_{li}^{j}$ is independent of $\hat{\vect{h}}_{li}^j$ and has correlation matrix $
\vect{C}_{li}^{j} = \mathbb{E} \{ \tilde{\vect{h}}_{li}^{j} ( \tilde{\vect{h}}_{li}^{j} )^{\Htran} \}  = \vect{R}_{li}^{j} - \vect{\Phi}_{li}^j
$ with $\vect{\Phi}_{li}^j =\vect{R}_{li}^{j}
{(\Psiv_{li}^{j})}^{-1}  \vect{R}_{li}^{j}$.
\end{theorem}

\subsection{Spectral Efficiency}
An \emph{achievable SE} is any number that is below the capacity. While the classical ``Shannon formula'' cannot be applied when the receiver has imperfect CSI, there exist well-established capacity lower bounds that can be used \cite{EmilNowPub17,Marzetta2016a}.

\begin{theorem}\label{theorem:uplink-SE}
If MMSE channel estimation is used, an UL SE of UE $k$ in cell $m$ is
\begin{equation} \label{eq:uplink-rate-expression-general}
\begin{split}
\mathsf{SE}_{k,m} = \frac{\tau_u}{\tau_c} \mathbb{E} \left\{ \log_2  \left( 1 + \gamma_{k,m}  \right) \right\} \quad \textrm{[bit/s/Hz] }
\end{split}
\end{equation}
with ${\tau_u}/{\tau_c}$ accounting for the fraction of samples used for UL data and\begin{align}
\!\!\!\gamma_{k,m} 
\!=\! \frac{ |  \vect{v}_{mk}^{\Htran} \hat{\vect{h}}_{mk}^m |^2  }{ 
 \vect{v}_{mk}^{\Htran} \! \left(   \sum\limits_{l=1,l\ne j}^{M}\sum\limits_{i=1}^K \hat{\vect{h}}_{li}^m {(\hat{\vect{h}}_{li}^m)}^{\Htran}\!\!\!+\!\!\!\!\!\sum\limits_{i=1,i\ne k}^K \!\!\!\hat{\vect{h}}_{mk}^m {(\hat{\vect{h}}_{mk}^m)}^{\Htran} \!\!\!+ \! \vect{Z}_m\!\!\right) \vect{v}_{mk}  
}   \label{eq:uplink-instant-SINR}
\end{align}
where $\vect{Z}_m= \sum_{l=1}^{M} \sum_{i=1}^{K} (\vect{R}_{li}^{m} - \vect{\Phi}_{li}^m) +  \frac{\sigma^2}{p}  \vect{I}_{M}$.
\end{theorem}
Two popular heuristic linear detectors are maximum-ratio (MR) combining
\begin{equation} \label{eq:MMSE-combining}
\!\!\vect{V}_{m}^{\rm{MR}} \!\triangleq \left[ \vect{v}_{j 1} \, \ldots \, \vect{v}_{j K}  \right] = \widehat{\vect{H}}_{m}^m
\end{equation}
with $\widehat{\vect{H}}_{j}^m = [\hat{\vect{h}}_{m1}^m, \ldots, \hat{\vect{h}}_{mK}^m]$ and zero-forcing (ZF) combining \cite{Marzetta2016a}. Instead of resorting to heuristics, we notice that the SINR in \eqref{eq:uplink-rate-expression-general}  has the form of a generalized Rayleigh quotient. Hence, the maximum is achieved by \cite{BHS18}:
\begin{equation} \label{eq:MMSE-combining}
\!\!\vect{V}_{m}^{\rm{M-MMSE}} \!\triangleq \left[ \vect{v}_{m 1}, \ldots, \vect{v}_{m K}  \right]  \! =\!\! \Bigg( \! \sum\limits_{l=1}^{M} \widehat{\vect{H}}_{l}^m {(\widehat{\vect{H}}_{l}^m)}^{\Htran} \!\!\!+ \vect{Z}_m  \!\Bigg)^{\!-1}  \!\!\!\!\widehat{\vect{H}}_{m}^m.
\end{equation}
This optimal combining scheme was introduced in \cite{EmilEURASIP17,BHS18} and called multicell MMSE (M-MMSE) combining. The ``multicell'' notion was used to differentiate it from the single-cell MMSE (S-MMSE) combining scheme \cite{Hoydis2013a}, which is widely used in the Massive MIMO literature. Next, we consider both MR and M-MMSE combining.

\subsection{Problem Formulation}
Denoting by $\rho_{k,m}$ the binary variable taking value $1$ when user $k$ is served by \gls{bs} $m$ and zero otherwise, the system sum-rate is expressed as 
\beq\label{Eq:SR}
\text{SR}=B\frac{\tau_{u}}{\tau_{c}}\sum_{k=1}^{K}\sum_{m=1}^{M}\rho_{k,m}\log_{2}(1+\gamma_{k,m})
\eeq
with $B$ denoting the communication bandwidth. The aim of this work is to maximize the sum-rate in \eqref{Eq:SR} with respect to the assignment variables ${{\boldsymbol {\rho}}}=\{\rho_{k,m}\}_{k,m}$. This leads to the optimization problem:
\begin{subequations}\label{Prob:UserBS}
\begin{align}
&\!\!\!\!\ds\max_{{{\boldsymbol {\rho}}}}\sum_{k=1}^{K}\sum_{m=1}^{M}\rho_{k,m}\log_{2}(1+\gamma_{k,m})\label{Prob:UserBSa}\\
&\!\!\!\!\;\textrm{s.t.}\;\sum_{m=1}^{M}\rho_{k,m}\leq 1\;,\forall\;k=1,\ldots,K\label{Prob:UserBSb}\\
&\!\!\!\!\;\;\quad\;\sum_{k=1}^{K}\rho_{k,m}\leq d_{m},\;\forall\;m=1,\ldots,M\label{Prob:UserBSc}\\
&\!\!\!\!\;\;\quad\; \rho_{k,m}\in\{0,1\}\;,\;\forall\;m=1,\ldots,M,\;\forall\;k=1,\ldots,K\label{Prob:UserBSe}
\end{align}
\end{subequations}
wherein \eqref{Prob:UserBSb} and \eqref{Prob:UserBSc} ensure that each user can be associated to only one \gls{bs} and that each \gls{bs} can serve at most $d_{m}$ users, while \eqref{Prob:UserBSe} is due to the integrality of the association variables. In particular, \eqref{Prob:UserBSc} can be used to enforce specific load balancing policies, constraining the maximum number of users that can be served by each \gls{bs}. Of course, it should be ensured that $\sum_{m=1}^{M}d_{m}\geq KM$ in order for \eqref{Prob:UserBSc} to be feasible. 

Problem \eqref{Prob:UserBS} can be seen as a linear integer program, for which a general solution method is the branch-and-cut algorithm, that requires solving a series of continuous relaxations of \eqref{Prob:UserBS}. Nevertheless, solving \eqref{Prob:UserBS} turns out to be computationally easier. Indeed, the facts that $\rho_{k,m}\in\{0,1\}$ for all $k$ and $m$, and that  $d_{m}$ is an integer number for all $m$, imply that the constraint matrix of  \eqref{Prob:UserBS} is totally uni-modular, and thus the global solution of \eqref{Prob:UserBS} coincides with the global solution of its continuous relaxation. Therefore, applying the relaxation $\rho_{k,m}\in[0,1]$ in \eqref{Prob:UserBS} leads to a continuous linear problem, which can be solved with polynomial complexity by standard linear programming, and whose solution is also the global solution to \eqref{Prob:UserBS}. However, although enjoying polynomial complexity, the number of variables in \eqref{Prob:UserBS} amounts to $KM^{2}$, which makes it challenging to solve \eqref{Prob:UserBS} in real-time in typical Massive MIMO networks where a large number of users must be served. Moreover, the optimal association changes each  time at least one of the users changes its position. This requires to solve \eqref{Prob:UserBS} once again and before yet another change in the users' positions occurs. In order to address this issue, next section shows how \eqref{Prob:UserBS} can be solved with a complexity amenable to real-time user-cell association optimization by resorting to the use of deep \glspl{ann}.

\section{Sum rate maximization by deep learning}\label{Sec:Approach}
\begin{figure}\vspace{-0mm}
  \begin{center}
  \includegraphics[scale=0.5]{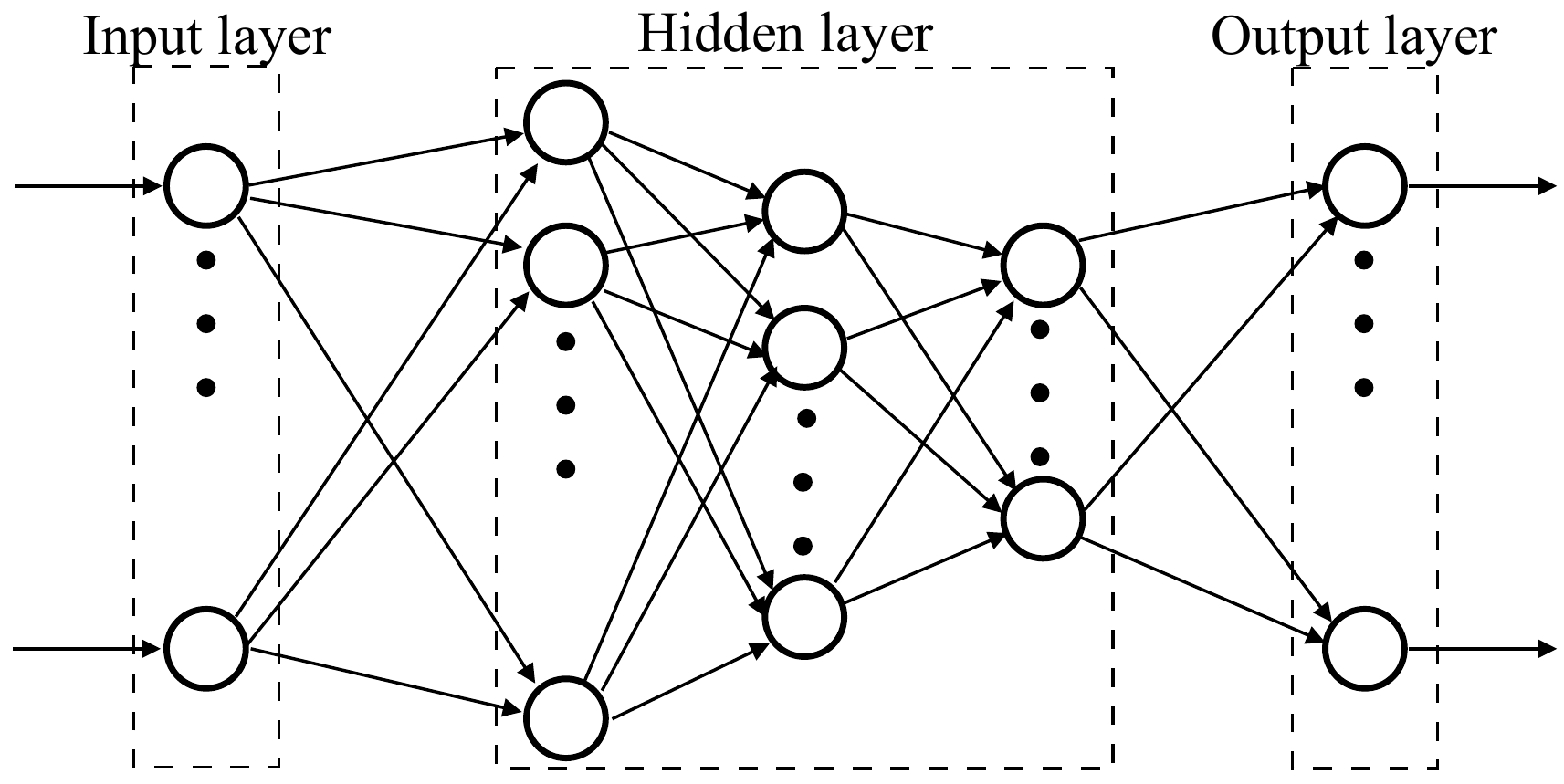}
  \caption{General scheme of a deep feedforward \gls{ann} with fully-connected layers.}\vspace{-0.7cm}
  \label{Fig:ANN_Scheme}
 \end{center}
\end{figure}

The proposed approach builds on the result that fully-connected feedforward \glspl{ann} are universal function approximators \cite{Hornik89,Bengio2016} and on the fact that the association problem in \eqref{Prob:UserBS} can be regarded as that of estimating an unknown map from the users positions in the service area, expressed as the coordinate pairs $\{(x_{k,m},y_{k,m})\}_{k,m}$, and problem parameters $\{d_{m}\}_{m}$, to the optimal association rule ${{\boldsymbol {\rho}}}^{*}\in\{0,1\}^{KM}$, namely
\beq\label{Eq:F}
{\cal F}:{\bf a}=\{x_{k,m},y_{k,m},d_{m}\}_{k,m}\in\mathbb{R}^{(2K+1)M}\to {{\boldsymbol {\rho}}}^{*}.
\eeq
Then, an \gls{ann} can be trained to learn the unknown map \eqref{Eq:F}, according to the architecture shown in Fig. \ref{Fig:ANN_Scheme}. 
Specifically, the input layer has $(2K+1)M$ neurons that forward the input vector ${\bf a}$ to the rest of the \gls{ann}, which is composed of $L$ hidden layers, and an output layer with $KM$ neurons providing an estimate ${{\boldsymbol {\rho}}}_{ANN}$ of ${{\boldsymbol {\rho}}}^{*}$. For all $\ell=1,\ldots,L+1$, the $\ell$-th layer is composed of $N_{\ell}$ neurons, with neuron $n$ computing 
\beq\label{Eq:TransferFunction}
{\bf z}_{\ell}(n)=f_{n,\ell}\left({\bf w}_{n,\ell}^{T}{\bf z}_{\ell-1}+b_{n,\ell}\right)
\eeq
wherein ${\bf z}_{\ell}$ denotes the $N_{\ell+1}\times 1$ output vector of Layer $\ell$, ${\bf w}_{n,\ell}\in\mathbb{R}^{N_{\ell-1}}$ and $b_{n,\ell}\in\mathbb{R}$ are neuron-dependent weights and bias terms, while $f_{n,\ell}$ is the activation function of neuron $n$ in layer $\ell$.

The goal is to configure the weights and bias terms of the \gls{ann} in such a way that the input-output relationship of the \gls{ann} emulates the map between ${\bf a}$ and ${{\boldsymbol {\rho}}}^{*}$, so that the \gls{ann} output ${{\boldsymbol {\rho}}}_{ANN}$ is a reliable estimate of ${{\boldsymbol {\rho}}}^{*}$. This can be achieved by training the \gls{ann} as described next.

\subsection{\gls{ann} training}
In order to successfully train the \gls{ann}, it is required to have a training set containing multiple pairs $\{({{\boldsymbol {\rho}}}_{nt}^{*},{\bf a}_{nt})\}_{nt=1}^{N_T}$, with ${{\boldsymbol {\rho}}}_{nt}^{*}$ being the optimal association when the input is ${\bf a}_{nt}$.  
Then, denoting by ${{\boldsymbol {\rho}}}_{nt}({\bf W},{\bf b})$ the actual \gls{ann} output corresponding to the training input ${\bf a}_{nt}$, with ${\bf W}=\{{\bf w}_{n,\ell}\}_{n,\ell}$ and ${\bf b}=\{b_{n,\ell}\}_{n,\ell}$, the training process consists of minimizing the loss between actual and desired output, namely solving:
\beq
\ds\min_{{\bf W},{\bf b}}\;\frac{1}{N_{T}}\sum_{nt=1}^{N_{T}}{\cal L}({{\boldsymbol {\rho}}}_{nt}({\bf W},{\bf b}),{{\boldsymbol {\rho}}}_{nt}^{*})
\eeq
with ${\cal L}(\cdot,\cdot)$ any suitable measure of the distance between the $nt$-th true and actual training sample, e.g. the mean squared error. This can be accomplished by means of gradient-based  descent algorithms, i.e. iteratively updating the parameters according to the formulas:
\begin{align}
{\bf W}(t+1)&={\bf W}(t)-\alpha\nabla L({\bf W}(t))\\
{\bf b}(t+1)&={\bf b}(t)-\alpha\nabla L({\bf b}(t))
\end{align}
with $\alpha$ the learning rate. 

Once the parameters ${\bf W}$ and ${\bf b}$ are configured, the \gls{ann} can estimate the optimal user-cell association also for new, i.e. not contained in the training set, input vectors ${\bf a}$. As a result, every time the positions of the users in the network change, the new association rule can be computed by simply feeding the new ${\bf a}$ to the \gls{ann}, without having to actually solve \eqref{Prob:UserBS}. As further elaborated next, this grants a huge complexity reduction and enables to update the user-cell association in real-time as the mobile users move around the coverage area. 

\subsection{Online implementation and complexity}
The complexity of the proposed method depends on two main tasks:
\begin{itemize}
\item[(a)] Use the trained \gls{ann} to compute the user-cell association.
\item[(b)] Building the training set and implementing the training procedure.
\end{itemize}
Among the two tasks above, computing the output of a trained \gls{ann} is by far the least complex. Indeed, once the \gls{ann} is trained, the output is essentially obtained by computing $\sum_{\ell=1}^{L+1}N_{\ell-1}N_{\ell}$ real multiplications\footnote{The complexity related to additions is neglected as it is much smaller than that required for multiplications.} and evaluating $\sum_{\ell=1}^{L+1}N_{\ell}$ activation functions. Essentially, it can be seen that the trained \gls{ann} provides a closed-form expression that relates the user-cell association to the users' positions in the area to serve. Moreover, it is worth-observing that, thanks to the fact that the proposed \gls{ann} operates based on the users positions, it does not require to compute the $KM^{2}$ rates of each user towards every possible \gls{bs}. Instead, only the $2KM$ real numbers defining the users' position coordinates are required. 

The training algorithm requires implementing a gradient-based algorithm, which might be computationally intensive for large training set sizes. However, the complexity is greatly diminished by the use of stochastic gradient descent algorithms, which evaluate the gradients over random subsets of the complete training set, called \emph{mini-batches} \cite[Ch. 8]{Bengio2016}, and leveraging  the \emph{back-propagation} algorithm \cite[Ch. 6.5]{Bengio2016}, which greatly speeds up gradient computations, enabling parallel computing, and ultimately making the complexity of implementing the training algorithm well affordable. 

On the other hand, most of the complexity of the training phase lies in the task of building the training set. Indeed, in order to build the training set, it is necessary to actually solve Problem \eqref{Prob:UserBS} for many different realizations of the users positions ${\bf a}$, which requires using traditional optimization theory. At a first sight, this might seem to defeat the purpose of using the proposed \gls{ann}-based approach, but this is not the case for two main reasons:
\begin{itemize}
\item The training set can be generated \emph{off-line}. Thus, a much higher complexity can be afforded and real-time constraints do not apply.
\item The training set can be updated at a much longer time-scale than the rate at which the users' positions in the network vary. In other words, the training set can be updated at a much longer time-scale than that at which Problem \eqref{Prob:UserBS} should be solved if traditional resource allocation approaches were used.
\end{itemize}
These considerations support the claim that the proposed \gls{ann}-based approach grants a huge complexity reduction over traditional approaches, enabling to update the users-cell association following the users' mobility in the service area. 

\section{Numerical analysis}
The performance of the proposed \gls{ann} has been numerically analyzed considering the uplink of a Massive MIMO system wherein $4$ \glspl{bs} are deployed in a square area with edge $1\,\textrm{km}$ at points with coordinates $(250,250)\,\textrm{m}$, $(250,750)\,\textrm{m}$, $(750,250)\,\textrm{m}$, $(750,750)\,\textrm{m}$, serving $40$ users randomly placed in the coverage area. Each \gls{bs} is equipped with $N_{R}=64$ antennas, while all mobile users have a single antenna. A uniform uplink power $p$ of $20\,\textrm{dBm}$ is considered for all users, while a common receive noise power $\sigma^2$ of $-94\,\textrm{dBm}$ is assumed for all \glspl{bs}. The communication bandwidth is $20\,\textrm{MHz}$ and  the propagation channels follow the local scattering model \cite{EmilNowPub17}. 

A training set of $N_{T}=155000$ samples has been generated considering independent realizations of the users' positions in the service area, and solving the corresponding instance of Problem \eqref{Prob:UserBS}, with $d_{m}=15$ for all $m$. 
The training will be provided upon request while the Matlab code available online at \href{https://github.com/lucasanguinetti/}{https://github.com/lucasanguinetti/}
allows to generate further samples.
In particular, the 140000 data samples have used as training set, while the remaining 15000 samples have been used as validation set for hyperparameter tuning. A feedforward \gls{ann} has been used, having $L=3$ fully connected layers with $128$, $64$, $64$ neurons, respectively, plus an output layer with $KM=40$ neurons. Layers $1$ and $3$ have a ReLU activation function, while Layer $2$ and the output layer have a sigmoidal activation function. The ADAM training algorithm with Nesterov's momentum has been employed for training \cite{Kingma2015_ADAM}, using the mean squared error as loss function i.e.:
\beq
{\text{MSE}}=\frac{1}{N_{TR}}\sum_{nt=1}^{N_{T}}\left\|{{\boldsymbol {\rho}}}_{nt}-{{\boldsymbol {\rho}}}_{nt}^{*}\right\|^{2}\;.
\eeq 
The training algorithm has been implemented using the open source python library Keras \href{https://keras.io}{https://keras.io}, setting the number of training epochs to $50$. The code is available online at \href{https://github.com/lucasanguinetti/}{https://github.com/lucasanguinetti/}.
The MSE obtained over the training and validation sets are reported in Tab. \ref{Tab:MSE} versus the epoch number. It is seen that the ANN neither underfits nor overfits the training data. 
\begin{table}[!h]\label{Tab:MSE}
\centering
\begin{tabular}{|c|c|c|}
\hline
 & Training MSE & Validation MSE \\
 \hline
Epoch 1 & $0.0116$ & 0.0113 \\
\hline
Epoch 5 & $0.0100$ & 0.0116 \\
\hline
Epoch 10 & $0.0093$ & 0.0104 \\
\hline
Epoch 15 & $0.0091$ & 0.0096 \\
\hline
Epoch 20 & $0.0090$ & 0.0091 \\
\hline
Epoch 25 & $0.0089$ & 0.0089 \\
\hline
Epoch 30 & $0.0087$ & 0.0092 \\
\hline
Epoch 35 & $0.0085$ & 0.0087\\
\hline
Epoch 40 & $0.0083$ & 0.0089 \\
\hline
Epoch 45 & $0.0082$ & 0.0087 \\
\hline
Epoch 50 & $0.0081$ & 0.0090\\
\hline
\end{tabular}
\end{table}
 
After training and validation, the performance of the resulting ANN has been evaluated over a test set of $15000$ data samples that has been independently generated from the training and validation sets. For each test sample, denoting by ${{\boldsymbol {\rho}}}_{ANN}=\{\rho_{k,m}\}_{k,m}$ the \gls{ann} output, user $k$ has been associated to \gls{bs} $\bar{m}$ if $\bar{m}=\text{arg}\max_{m}\;\rho_{k,m}$, and then the resulting sum-rate performance has been compared to the optimal solution of Problem \eqref{Prob:UserBS}.

\begin{figure}[!h]
\centering
\includegraphics[scale=0.33]{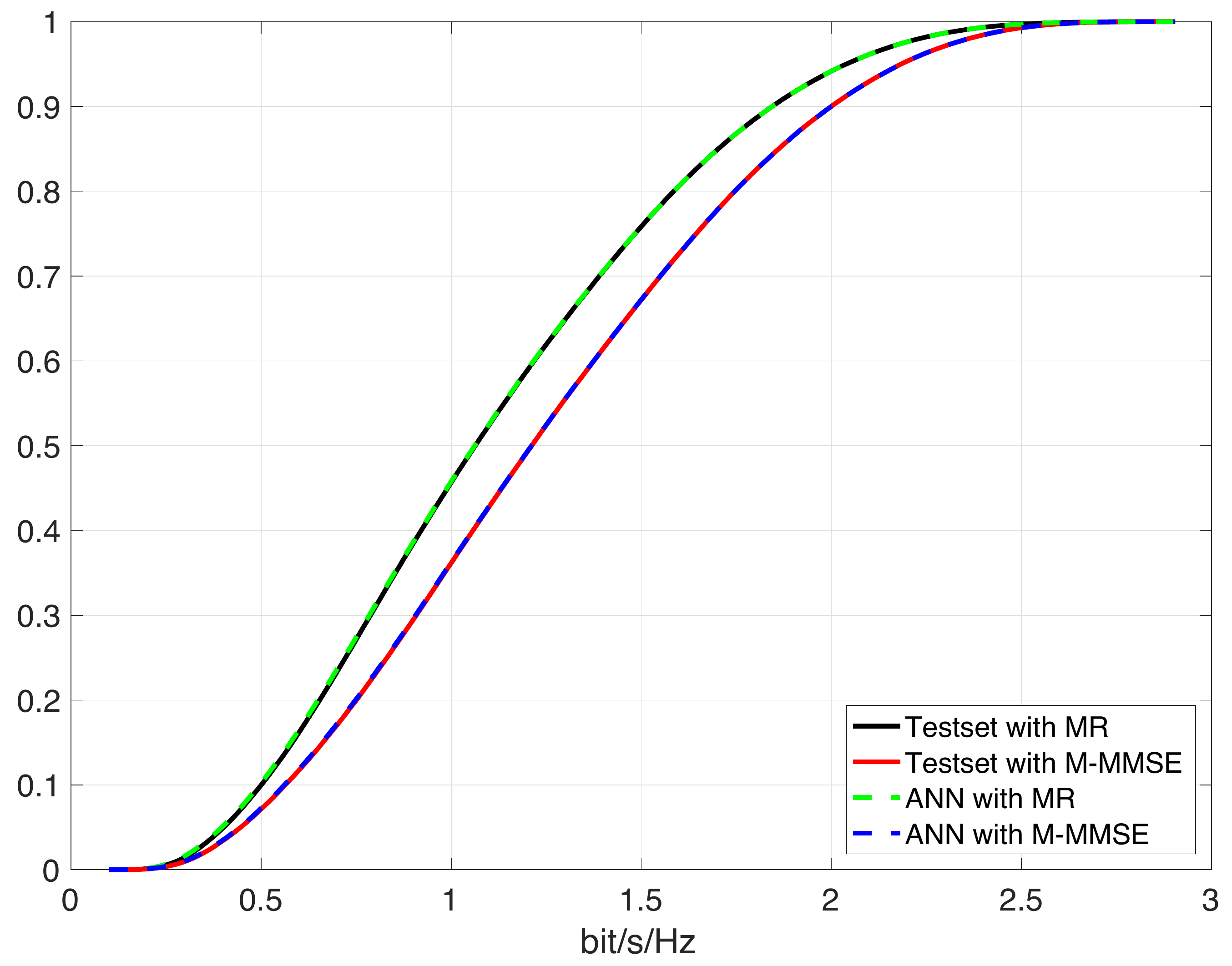}
\caption{CDF of the average users' rate over the testset for the \gls{ann}-based approach and the optimal allocation, with MMSE and MR reception.} \label{Fig:CDF}
\end{figure}

Fig. \ref{Fig:CDF} shows the cumulative distribution function (CDF) of the average users' rate over the test set for the following schemes:
\begin{itemize}
\item \gls{ann}-based association with MMSE reception.
\item Optimal association with MMSE reception.
\item \gls{ann}-based association with MR reception.
\item Optimal association with MR reception.
\end{itemize}
It is seen that in all cases the proposed \gls{ann}-based method is able to achieve virtually optimal performance with MMSE reception granting better performance than MR reception, as expected. 

\begin{figure}[!h]
\centering
\includegraphics[scale=0.33]{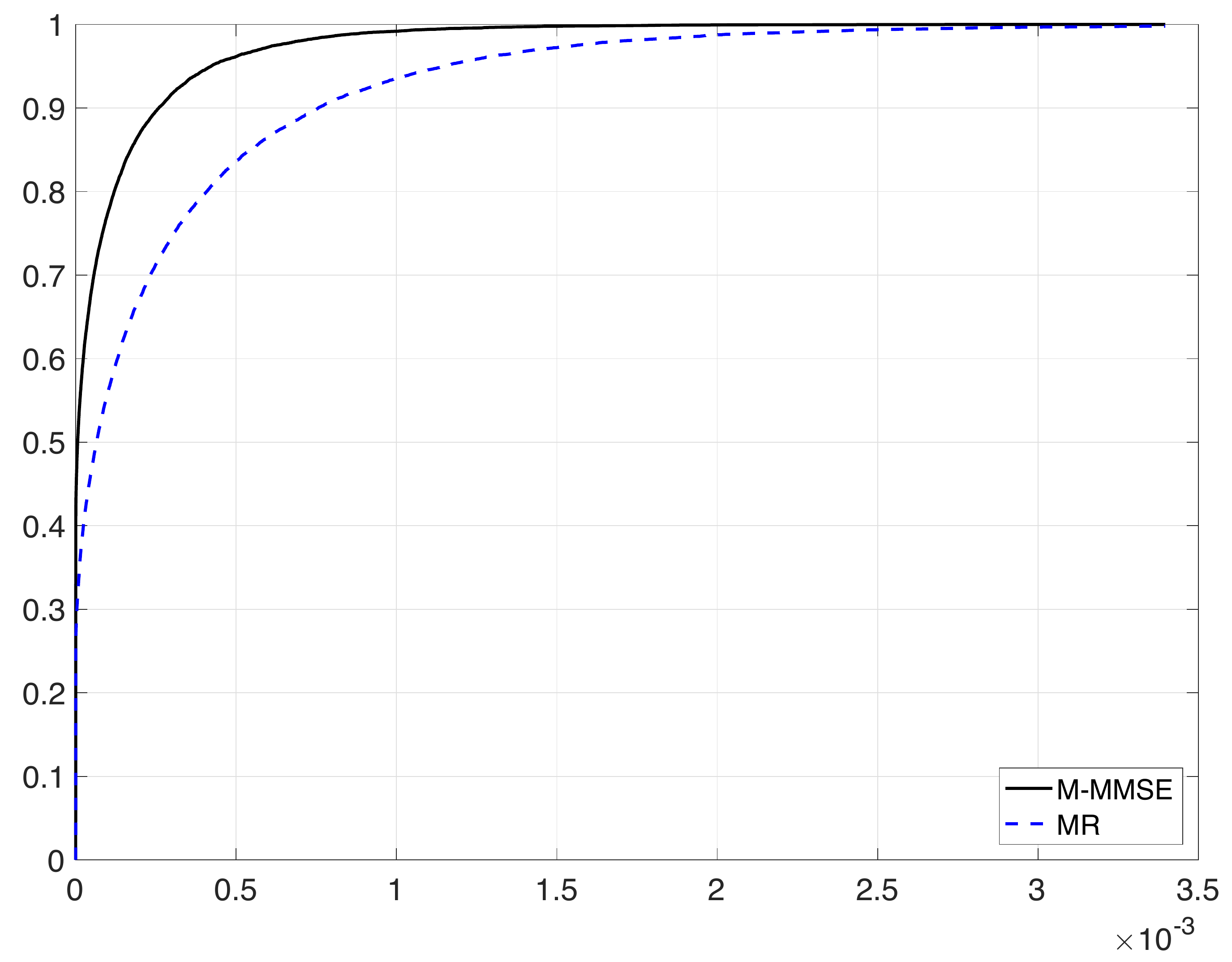}
\caption{CDF of the MSE over the testset for M-MMSE and MR reception.} \label{Fig:CDF_Err}
\end{figure}

Fig. \ref{Fig:CDF_Err} considers a similar scenario as Fig. \ref{Fig:CDF}, but shows the CDF of the mean squared error over the test set. Interestingly, it is observed that the error when M-MMSE reception is used is lower than when MR reception is employed, which implies that the \gls{ann}-based association performs better when the most performing reception scheme is used. This fact can be intuitively explained observing that the use of the M-MMSE receiver grants a better interference suppression than MR reception. In turn, this tends to decouple the allocation problem over the users and \glspl{bs}, thus simplifying the structure of the optimal allocation which can be more easily learnt by the \gls{ann}.

\section{Conclusions}\label{Sec:Conclusions}
This work investigated the use of deep learning for sum-rate maximization by user-cell association in Massive MIMO networks. Leveraging the universal approximation property of feedforward \glspl{ann}, it showed that a relatively small \gls{ann} is able to learn the optimal user-cell assignment based only on the users' positions in the service area. The use of an \gls{ann} grants huge complexity reductions in the assignment procedure compared with traditional approaches, thereby enabling to update the association rule in real-time, tracking the users' mobility pattern. 

\bibliographystyle{IEEEbib}
\bibliography{DeepLearning,FracProg,references}

\end{document}